# Toward Automated Detection of Microbleeds with Anatomical Scale Localization: A Complete Clinical Diagnosis Support Using Deep Learning

Jun-Ho Kim[1], Young Noh[2,3], Haejoon Lee[4], Seul Lee[1], Woo-Ram Kim[2], Koung Mi Kang[5,6], Eung Yeop Kim[7], Mohammed A. Al-masni[8,*], Dong-Hyun Kim[1,*]

[1] Department of Electrical and Electronic Engineering, College of Engineering, Yonsei University, Seoul, Republic of Korea
[2] Neuroscience Research Institute, Gachon University, Incheon, Republic of Korea
[3] Department of Neurology, Gachon University College of Medicine, Gil Medical Center, Incheon, Republic of Korea
[4] Department of Electrical and Computer Engineering, Carnegie Mellon University, Pittsburgh, Pennsylvania, USA
[5] Department of Radiology, Seoul National University Hospital, Seoul, Republic of Korea
[6] Department of Radiology, Seoul National University College of Medicine, Seoul, Republic of Korea
[7] Department of Radiology, Gachon University College of Medicine, Gil Medical Center, Incheon, Republic of Korea
[8] Department of Artificial Intelligence, College of Software & Convergence Technology, Daeyang AI Center, Sejong University, Seoul 05006, Republic of Korea

**Abstract**

Cerebral Microbleeds (CMBs) are chronic deposits of small blood products in the brain tissues, which have explicit relation to various cerebrovascular diseases depending on their anatomical location, including cognitive decline, intracerebral hemorrhage, and cerebral infarction. However, manual detection of CMBs is a time-consuming and error-prone process because of their sparse and tiny structural properties. The detection of CMBs is commonly affected by the presence of many CMB mimics that cause a high false-positive rate (FPR), such as calcification and pial vessels. This paper proposes a novel 3D deep learning framework that does not only detect CMBs but also inform their anatomical location in the brain (i.e., lobar, deep, and infratentorial regions). For the CMB detection task, we propose a single end-to-end model by leveraging the U-Net as a backbone with Region Proposal Network (RPN). To significantly reduce the FPs within the same single model, we develop a new scheme, containing Feature Fusion Module (FFM) that detects small candidates utilizing contextual information and Hard Sample Prototype Learning (HSPL) that mines CMB mimics and generates additional loss term called concentration loss using Convolutional Prototype Learning (CPL). For the anatomical localization task, we exploit the U-Net segmentation network to segment the brain anatomical structures. This task does not only tell to which region the CMBs belong but also eliminate some FPs from the detection task by utilizing anatomical information. We utilize Susceptibility-Weighted Imaging (SWI) and phase images as 3D input to efficiently capture 3D information. The results show that the proposed RPN that utilizes the FFM and HSPL outperforms the vanilla RPN and achieves a sensitivity of 94.66% vs. 93.33% and an average number of false positives per subject ($FP_{avg}$) of 0.86 vs. 14.73. Also, the anatomical localization task further improves the detection performance by reducing the $FP_{avg}$ to 0.56 while maintaining the sensitivity of 94.66%. The proposed CMB detection and anatomical location identification framework shows its feasibility as a complete clinical diagnosis support tool.

# 1 Introduction

Cerebral Microbleeds (CMBs) are chronic deposits of small blood products in the brain tissues and are generated due to damage to the vessel walls. CMBs usually occur close to the arteries and capillaries [1, 2]. Microbleeds are commonly detected in individuals of advancing age and patients with cerebrovascular disease [3]. Especially, Microbleeds are more prevalent in patients with Alzheimer's disease (AD), dementia, ischemic, and hemorrhagic stroke. Recently, CMBs are reported to be related to cognitive decline, intracerebral hemorrhage, cerebral infarction, and recurrence of transient ischemic attack [4, 5]. In addition, it is observed that CMBs are useful biomarkers for pathologic damage to small vessels from hypertension or cerebral amyloid angiopathy (CAA) [6, 7]. Not only the presence of CMBs, but also the location of CMBs (i.e., lobar, deep, and infratentorial regions) determines what disease they are associated with. For example, the CMBs located in the lobar region are closely related to CAA associated with Alzheimer's disease [8]. In the case of deep CMBs, they are related to lacunar infarcts, hypertension, and diastolic blood pressure variability. The CMBs in the infratentorial region are associated with systolic blood pressure variability and migraine without aura [9-12]. In terms of cognitive decline, the lobar CMBs are related to general cognitive function, executive function, memory, and process speed, while the deep and infratentorial CMBs are associated with psychomotor speed and attention [4].

[*] Corresponding author

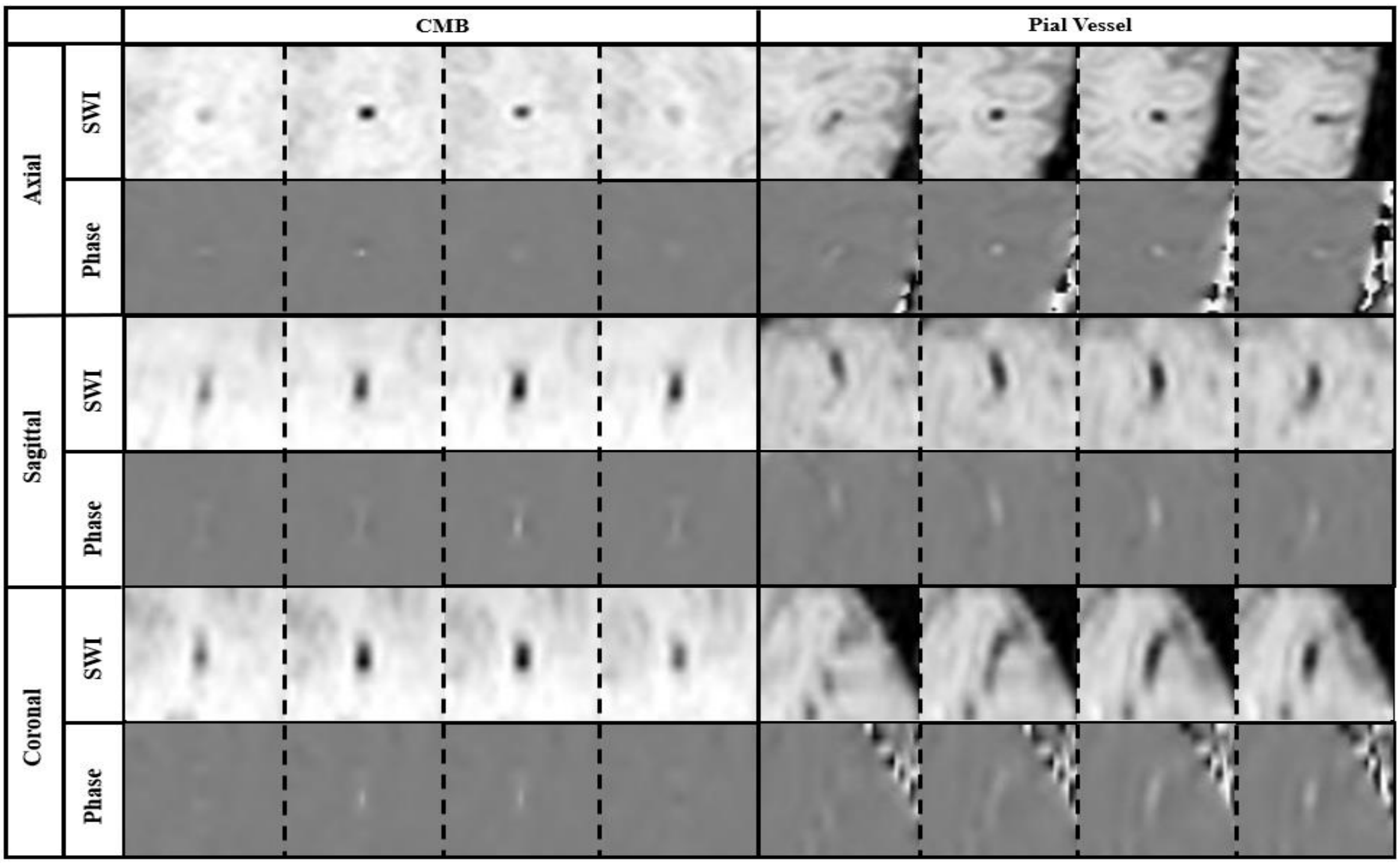

Fig. 1. Comparison between CMB and pial vessel through consecutive slices in the axial, sagittal, and coronal planes. The scan resolution of these MR images is 0.5×0.5×2 $mm^2$.

Magnetic Resonance Imaging (MRI) is the most widely utilized modality for CMB detection. Vessel bleeding with a small diameter of 200 um can be screened utilizing the Susceptibility-Weighted Images (SWI) generated from Gradient-Recalled Echo (GRE) MRI pulse sequences [13]. However, there are challenging factors in CMB detection. CMBs, which are round or elliptical lesions with a size of 2-10 mm, are very sparse and small compared to the whole brain tissue [14, 15]. Another challenging factor is the presence of many CMB mimics that appear with hypointensities similar to CMBs in SWI images (e.g., calcification and pial vessels). It is observed that calcification can be distinguished from CMBs using the phase images since they have opposite intensities. However, in the case of a pial vessel, it is difficult to be distinguished from CMBs because they have the same intensity in both SWI and phase images as shown in Figure 1. Therefore, they are discriminated by inspecting the consecutive slices or views of multiple directions (e.g., coronal and sagittal) [14]. For these reasons, manual detection and anatomical localization are time-consuming, laborious, and the inspection results are subjective among neuroradiologists. These problems can be alleviated using an automated detector as an auxiliary tool, which assists in increasing the time-efficiency of microbleeds detection.

Early works in CMB detection mainly employed hand-crafted features to distinguish CMBs from CMB mimics. For example, Barnes et al. proposed a semi-automated method based on the statistical thresholding and Support Vector Machine (SVM) classifier to identify CMBs [16]. Kuijf et al. utilized the 3D Fast Radial Symmetry Transform (FRST) to detect CMB candidates [17]. Ghafaryasal et al. proposed a computer-aided system using simple intensities, size, and shape features with local image descriptors [18]. Fazlollahi et al. proposed a cascaded model utilizing cascaded random forest classifiers trained on radon-based features [19].

The main drawback of these conventional methods is that it is difficult to extract the effective features that differentiate between CMBs and CMB mimics effectively. This can be solved by applying a deep learning Convolutional Neural Network (CNN). CNN models have shown great progress not only in the computer vision field but also in medical image analysis tasks, including the detection of cerebral small vessel diseases [20-26].

In the literature, there were two types of CMB detection methods using deep learning CNN (i.e., single-stage detector and two-stage detector). The single-stage detector detects the CMBs directly with only one deep learning model. For example, Hong et al. employed the 2D-ResNet-50 to distinguish between CMBs and CMB mimics, recording sensitivity of 95.71% and false positives per subject ($FP_{avg}$) of 3.4 [27]. Similarly, Wang et al. utilized the 2D-DenseNet to classify CMB or non-CMB samples and achieved a sensitivity of 97.78% and an $FP_{avg}$ of 11.8 [28]. These single-stage detectors chose a classification model to perform better performance than the detection model. However, unlike the detection model that provided multiple outputs by classifying all regions of the input image at once, the classification model only classified the central object of the input image. For this reason, it is necessary to inference several times while moving the center of the cropped patch step by step, which resulted in a high computational cost and long execution time. Recently, Lee et al utilized the detection models that are trained in parallel for three planes (axial, coronal, and sagittal) and achieved a sensitivity of 93.33 and $FP_{avg}$ of 1.52 [29]. However, it is not a true single-stage detector because three detection models must be trained separately.

On the other hand, the two-stage detector approach consisted of two sequential models (detection and classification) The first stage is usually used for screening (i.e., potential candidate detection), while the second stage is responsible to distinguish true CMBs from CMB mimics (i.e., false positives (FPs) reduction). For example, Dou et al. utilized a 3D fully convolutional network (3D-FCN) for the first stage and 3D-CNN for the second stage, and the detector achieved an overall sensitivity of 91.45% and $FP_{avg}$ of 2.74 [23]. Liu et al. exploited a 3D-FRST as the first stage and a 3D residual network (3D-ResNet) as the second stage, and the cascaded detector obtained an overall sensitivity of 95.24% and $FP_{avg}$ of 1.6 [24]. Similarly, Chen et al. adopted 2D-FRST for the first stage and 3D-ResNet for the second stage and reported an overall sensitivity of 81.9% and $FP_{avg}$ of 11.58 [25]. Al-masni et al. utilized YOLO-v2 for the first stage and 3D-CNN for the second stage, and the cascaded detector achieved an overall sensitivity of 88.3% and $FP_{avg}$ of 1.42 [26]. Sundaresana et al. adopted U-Net for the first stage and 3D-CNN for the second stage utilizing the knowledge distillation method, and the framework records an overall sensitivity of 93% and $FP_{avg}$ of 1.5 [30]. Unlike the single-stage approach, the two-stage detector required a relatively less computational cost since only the CMB candidates from the first stage are passed through the second stage classification model. In addition, most two-stage detectors outperformed the single-stage detectors; especially regarding the false-positive cases [23, 24, 26-28, 30]. However, the two-stage detector is not trained in an end-to-end manner, so there is a disadvantage that the second stage depends on how the first stage performs; especially regarding the lost false-negative cases. Recently, the performance of a single-stage detector utilizing only a detection model (i.e., YOLO-v2) has been improved by adding a post-processing step called Cerebrospinal Fluid (CSF) filtering; however, the performance was still much lower than the two-stage detector (i.e., YOLO-v2 + 3D-CNN) [31].

In this paper, we propose a single-stage 3D deep learning detection model for automatic CMBs detection. The proposed work utilizes both the SWI and phase images as 3D input to efficiently capture 3D information. The main contributions of this work are summarized as follows. First, we establish a new 3D CMBs detector by integrating a U-Net [32] as the backbone and Region Proposal Network (RPN) of Faster R-CNN in a single end-to-end network [33]. Second, we incorporate the Feature Fusion Module (FFM) into the model so that it can efficiently learn contextual information. Third, we develop a single-stage detector without the need for the classification model, which outperforms the two-stage detectors, by adding the newly proposed additional loss term from Hard Sample Prototype Learning (HSPL) inspired by Convolutional Prototype Learning (CPL) [34].

In addition, we extend our single-stage detector initially proposed in MICCAI2022 to a framework that not only detects CMBs but also informs their anatomical location by segmenting the brain structures [35]. Our new contributions of this work include: First, to the best of our knowledge, we develop an unprecedented location identification task that informs the anatomical location of CMBs (i.e., lobar, deep, and infratentorial regions) and further reduces false positives from detector. Second, we investigate the clinical feasibility of the proposed framework by evaluating the generalization capability on different unseen clinical data. Third, we compared the performance of our detector with the state-of-the-art method on our dataset.

The rest of this paper is structured as follows. Section 2 explains in detail the proposed deep learning framework that is composed of single-stage detector and location identification tool. Experimental results of the framework are presented and discussed in Section 3. Finally, we draw the conclusions in Section 4.

# 2 Materials and Methods

## 2.1 Datasets

We retrospectively collected brain MR images of patients with CMBs from Gachon University Gil Medical Center (GMC). A total of 114 subjects including 365 CMBs were acquired. 23 subjects including 75 CMBs were randomly selected for the testing, and the remaining 91 subjects including 290 CMBs composed the training dataset. The size range of CMBs is almost under 5 mm and its size limit reaches 10 mm. The subjects consisted of 59 patients with cognitively normal, seven patients with mild cognitive impairment, and 48 patients with dementia (e.g., Alzheimer's dementia, frontotemporal dementia, and traumatic brain injury). All subjects were scanned using 3T Verio and Skyra Siemens MRI scanners with the following imaging parameters: echo time (TE): 20 ms; repetition time (TR): 27 ms; flip angle (FA): 15°; bandwidth/pixel (BW/pixel): 120 Hz/pixel; resolution: 0.50×0.50 $mm^2$; slice thickness: 2 mm; and matrix size: 512×448×72.

Additionally, we collected a total of 94 subjects including 311 CMBs for generalization assessment. This testing data were collected from Seoul National University Hospital (SNUH) and were scanned using a 3T Biograph mMR Siemens MRI scanner with the following imaging parameters: TE: 20 ms; TR: 28 ms; FA: 15°; BW/pixel: 170 Hz/pixel; resolution: 0.4911×0.4911 $mm^2$; slice thickness: 3 mm; and matrix size: 448×392×52.

The data acquisition was performed under the relevant regulations and guidelines. The study was approved by the Institutional Review Board of both sites.

### 2.2 Labeling

To obtain labels for the detection task, the label of GMC dataset were labeled based on the agreement of all the raters (a neuroradiologist with 26 years of experience, a medical imaging researcher with eight years of experience, and a neurologist with 17 years of experience). The label of SNUH dataset were annotated by a neuroradiologist with 13 years of experience. The raters provided the exact center coordinates of CMBs' location of each 3D MRI subject using SWI and phase images.

For the anatomical localization task, the location of CMBs is categorized into three distinct regions based on the Brain Observer MicroBleed Scale (BOMBS): lobar [cortex; subcortical white matter], deep [thalamus; internal/external capsule; basal ganglia], and infratentorial [cerebellum; brain stem] [36]. The brain regions that are not included within these three categories or the regions where CMBs can never exist are labeled as 'none'. To generate the labels for this task, we utilized FreeSurfer developed based on T1 image, which represents the structure of the brain [37]. For this reason, we utilized T1-Magnetization Prepared Rapid Gradient Echo (T1-MPRAGE) images that are registered with SWI images. The FreeSurfer offers segmentation of brain based on several atlases. Among different atlases, we selected the Automatic subcortical segmentation (Aseg) atlas, which divides the brain into 27 subregions [38]. In these 27 subregions, the cerebral cortex, cerebral white matter, white matter hypointensities, corpus callosum posterior, corpus callosum mid posterior, corpus callosum central, corpus callosum mid anterior, corpus callosum anterior, hippocampus, and amygdala are included in the lobar; the caudate, putamen, pallidum, thalamus, accumbens area, and ventral diencephalon are included in the deep; the brain stem, cerebellum white matter, and cerebellum cortex are included in the infratentorial; and the lateral ventricle, temporal horn of lateral ventricle, third ventricle, fourth ventricle, cerebrospinal fluid, choroid plexus, vessel, and optic chiasm are included in 'none' category. It is of note that the Aseg atlas of FreeSurfer does not provide the segmentation of internal/external capsules, which are regions relevant to the deep CMBs category; however, they are incorporated into the cerebral white matter of the lobar category. Thus, the raters manually separated the internal/external capsule from the cerebral white matter.

### 2.3 Data Preprocessing

In this work, we trained the proposed framework using both the SWI and phase images. Before training, the data were preprocessed. Firstly, we applied the brain extraction tool (BET) for brain skull stripping[39]. This enables to reduce the number of false positives that could be caused by the skull. In addition, since there is a variance in pixel intensity for each subject, all MR images of each subject were normalized using min-max normalization, which brings all voxel intensities in the range of 0 to 1. Moreover, we applied slice interpolation to increase the number of slices in the z-direction to 224 slices in the case of the CMB detection task. The reason for this interpolation is to improve the nominal resolution of the z-direction. However, in the case of the anatomical localization task, we did not apply slice interpolation in order to maintain the precise segmentation labels. Due to the GPU memory problem that can occur when the entire brain MR volumes are used as inputs, data cropping is essential for 3D network training. The training and testing data were prepared by cropping the whole MR image into 128×128×128 voxels for the detection task and 64×64×16 voxels for the anatomical localization task. Since cropping the entire brain using the sliding window method and feeding all crops into the network causes high computational cost, we randomly cropped the data and used the crops as training inputs. In the case of testing, we utilized a sliding window to infer all regions of each subject.

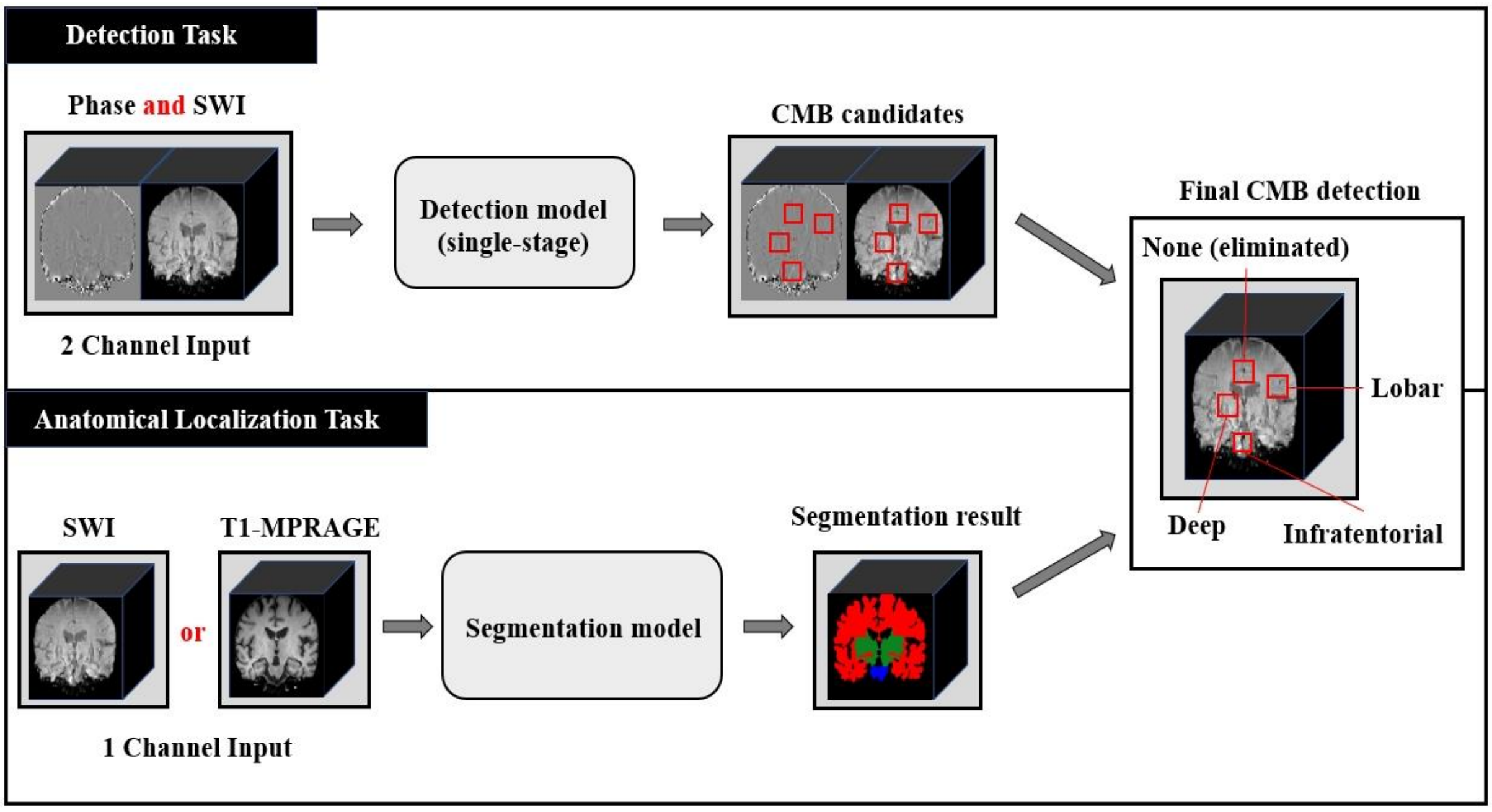

Fig. 2. Overview diagram of the proposed deep learning framework.

### 2.4 Proposed Deep Learning Framework

The proposed deep learning framework detects the CMBs first in the detection task and then informs the anatomical location (i.e., lobar, deep, and infratentorial regions) of all detected CMBs by segmenting the brain structures in the location identification task. The proposed detection task is an end-to-end single-stage detector that incorporates the U-Net and RPN of Faster R-CNN. For the location identification task, we utilized U-Net to segment the brain. Figure 2 illustrates the overall schematic diagram of the proposed framework.

#### 2.4.1 CMB Detection Task

The CMB mimics are considered the main challenge for previously developed methods since they cause the generation of many false positives. In this context, conventional methods utilized a classification model as a second stage to discriminate between true CMBs and CMB mimics. In contrast, our proposed single-stage detector eliminates the need for a second stage in the conventional works and instead incorporates the Feature Fusion Module (FFM) and Hard Sample Prototype Learning (HSPL).

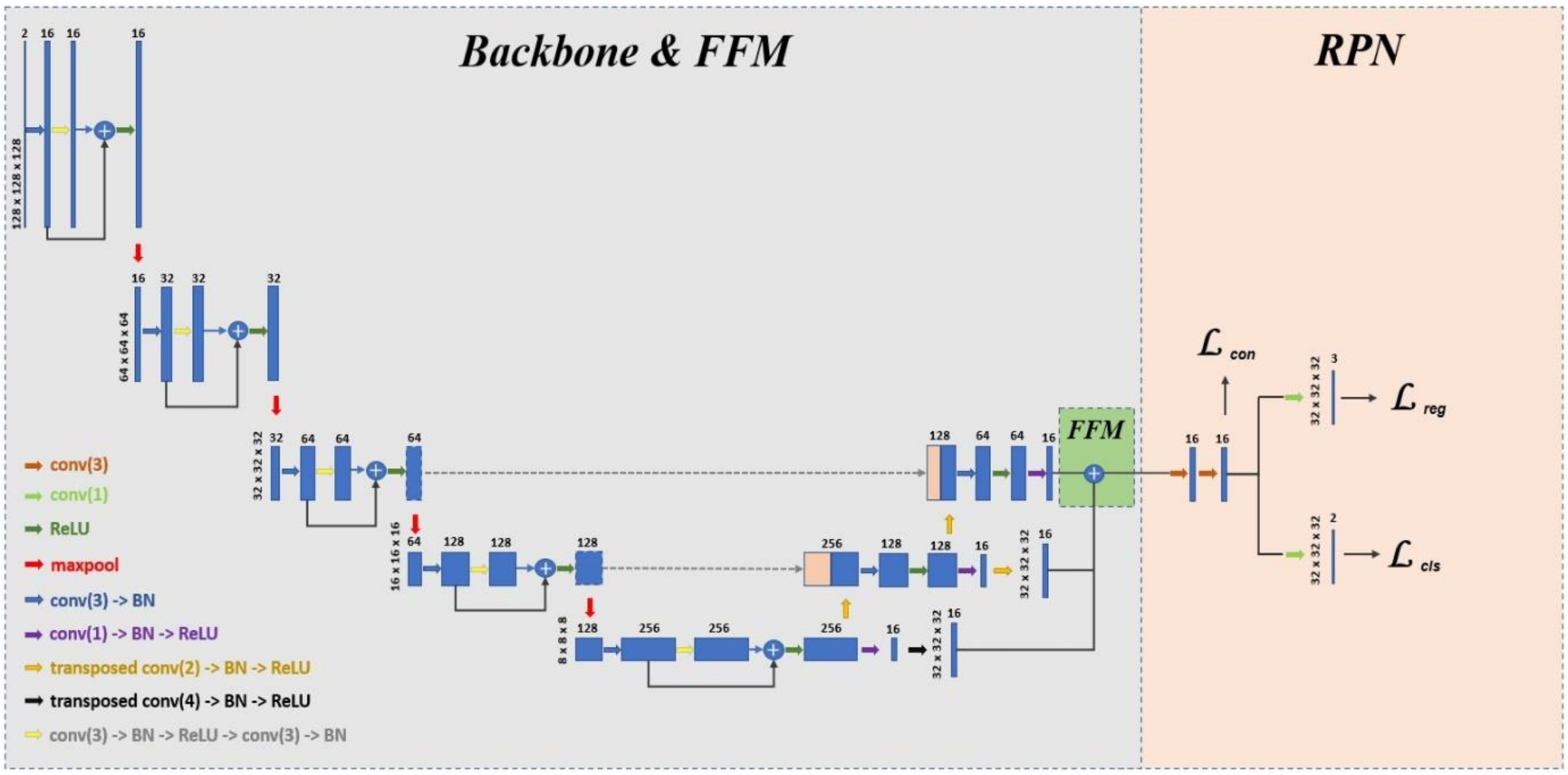

Fig. 3. The architecture of the detection task. The input data consists of SWI and phase images. conv(n): n×n×n convolutional layer, BN: batch normalization, transposed conv(n): n×n×n transposed convolutional layer, maxpool(n): n×n×n max pooling layer. The $L_{cls}$, $L_{reg}$, and $L_{con}$ are losses for classification, bounding box offset, and concentration learning, respectively.

#### 2.4.1.1 Feature Fusion Module (FFM)

The U-Net is structurally divided into two paths. The first path is a contracting path that captures the context features, while the other path is an expanding path that upsamples the reduced feature to its original resolution. The upsampled results are concatenated with feature maps of the same size in the contracting path. As shown in Figure 3, unlike the vanilla U-Net that expands the reduced feature map into original input size, we do not expand the reduced feature map into original input size because the bounding box regression module finds the fine-tuned center of bounding boxes. This method reduced the computational cost. In addition, we added a FFM that involves the contextual information into the final feature map [40, 41]. We utilized a 1×1×1 convolution layer to reduce the number of channels in each level of feature maps in the expanding path. The feature maps of the last two encoder levels are upsampled into the size of 32×32×32. After that, the final feature map of the backbone is generated by aggregating the upsampled feature maps with the feature map at the third level. This module makes the network more robust in distinguishing between CMBs and CMB mimics.

The RPN is a fully convolutional network, which receives the feature map from the backbone and probabilistically produces final region proposals. The number of final output channels is:

$$N_{out} = N_{cls} + N_{cd} \qquad (1)$$

Where $N_{cls}$ and $N_{cd}$ denote classes and dimension of the bounding box center coordinate. In this work, the size of bounding box is fixed to 20×20×20. Therefore, regression for the size of the bounding boxes and anchor boxes are not necessary. Since the input data is three-dimensional (x, y, z), the $N_{cd}$ is 3, and there are two classes, CMB and non-CMB, so the $N_{cls}$ is 2.

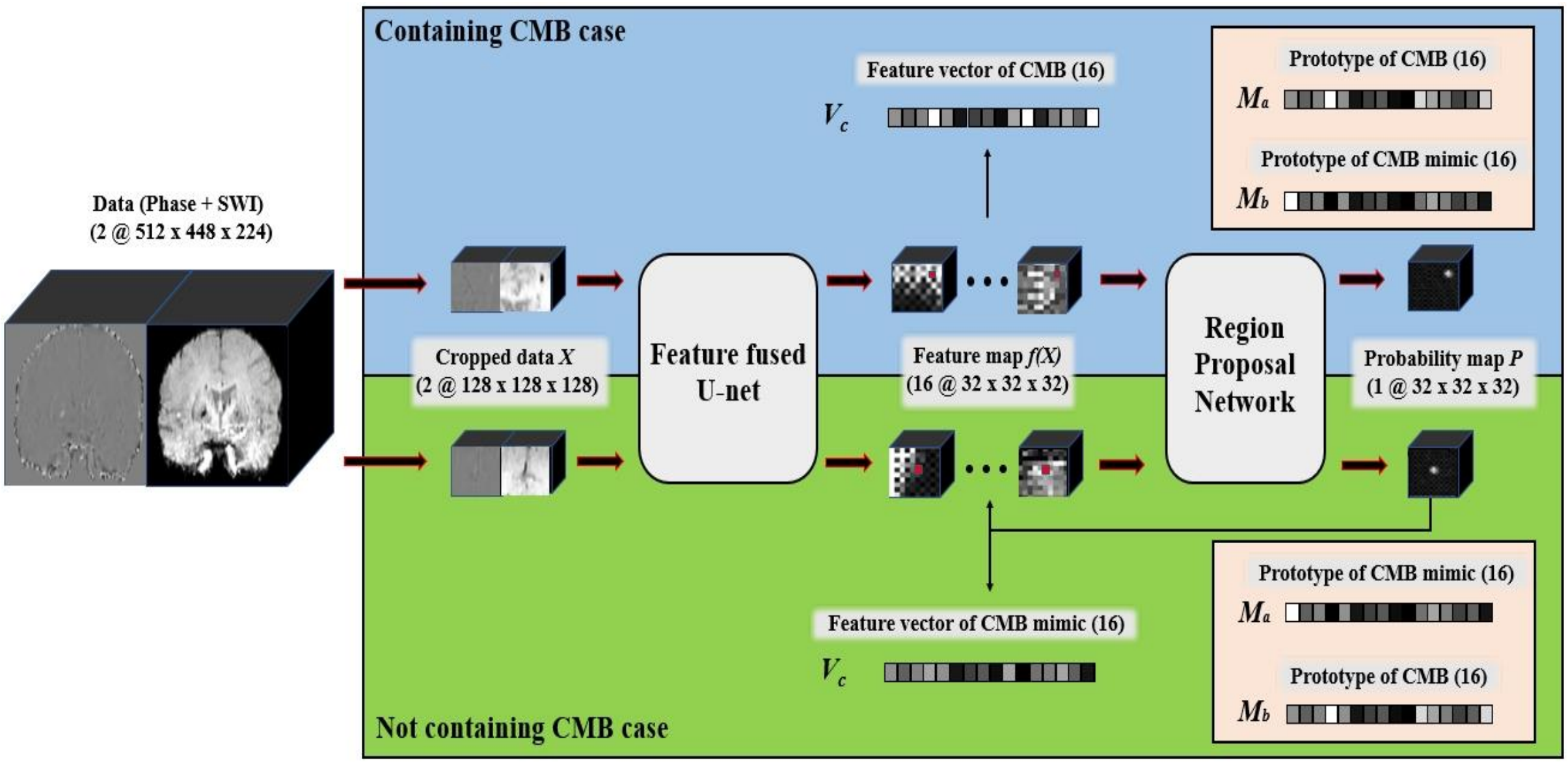

Fig. 4. Overview diagram of the proposed Hard Sample Prototype Learning (HSPL).

#### 2.4.1.2 Hard Sample Prototype Learning (HSPL)

The second stage of the conventional two-stage models can reduce the false positives because it was trained on a dataset consisting only of CMBs and CMB mimics (i.e., pure samples), excluding the background and easy samples. However, such two-stage approaches are not end-to-end learning and hence have the disadvantage that the second stage should be trained after finishing the training of the first stage. In other words, the performance of the second stage depends on how the first stage performs.

To enable our model to concentrate on CMB mimics without the need for employing a second stage, we developed a Hard Sample Prototype Learning (HSPL) that mines CMB mimics and generates concentration loss during training. Firstly, as illustrated in Figure 4, due to the sparse and tiny properties of CMBs, the HSPL crops the data based on the rule that the number of crops containing CMBs equals the crops that do not contain the CMBs. After the cropped data passes the backbone and RPN, the HSPL finds coordinates of CMB and CMB mimic using label and probability map $P \in \mathbb{R}^{d \times w \times h}$, where d, w, and h are depth, width, and height, respectively. In the case of data containing CMB, the coordinates of the CMB can be inferred directly from the ground-truth label. On the other hand, in the case of data not containing CMB, it is assumed that there is a CMB mimic in the cropped data, and the point with the highest confidence score in its probability map is considered as the point where the CMB mimic is located.

The final coordinate of CMB or CMB mimic is defined by:

$$c = \begin{cases} argmax_{i \in S}(P_i), & if\ X\ not\ contain\ CMB \\ i_l, & if\ X\ contain\ CMB \end{cases} \quad (2)$$

where $X$ represents the cropped data. $i$ and $i_l$ denote the coordinates of the highest probability point and the coordinates of CMB, respectively. Utilizing these coordinates, the HSPL creates a feature vector $V_c \in \mathbb{R}^{N_{ch}}$ by collecting the values located at coordinates $c$ in each channel of the feature map $f(X) \in \mathbb{R}^{N_{ch} \times d \times w \times h}$, where $N_{ch}$ indicates the number of feature map's channels. When the $V_c$ is extracted from the data containing CMB, the prototype of CMB becomes $M_a$, and the prototype of CMB mimic becomes $M_b$. Conversely, when the $V_c$ is extracted from the data not containing CMB, the prototype of CMB mimic becomes $M_a$, and the prototype of CMB becomes $M_b$. The $M_a$ and $M_b$ are trainable parameters, where they are automatically learned from the feature vectors $V_c$. The concentration loss is formulated as follows:

$$L_{con} = \frac{\|V_c - M_a\|_2^2 - \|V_c - M_b\|_2^2}{\|V_c - M_a\|_2^2 + \|V_c - M_b\|_2^2} + n \quad (3)$$

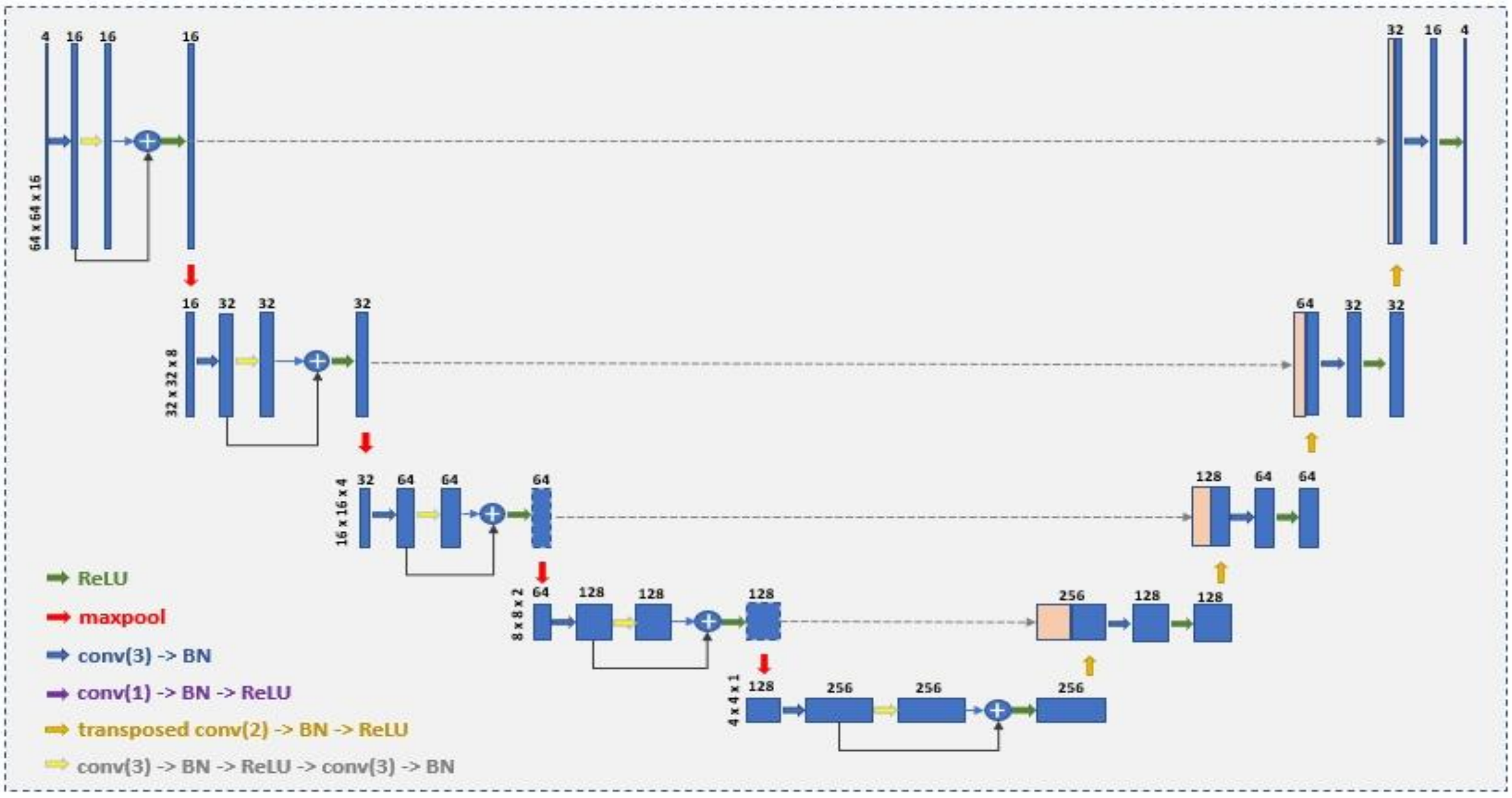

Fig. 5. The architecture of the anatomical location identification task. The input data consists of either a SWI or T1-MPRAGE image and automatically save absolute coordinate information for x, y, and z axes.

where n is the margin and set to 1 so that the concentration loss ranges from 0 to 2. The loss makes the distance between $V_c$ and $M_a$ closer and the distance between $V_c$ and $M_b$ farther in the feature space.

The final loss function of our network is computed as follows:

$$L_{final} = \lambda_1 L_{cls} + \lambda_2 L_{reg} + \lambda_3 L_{con} \quad (4)$$

where $L_{cls}$ is the classification loss implemented by focal loss [42], and $L_{reg}$ is regression loss following the bounding box regression of YOLO-v2 [43]. The $\lambda_1$, $\lambda_2$, and $\lambda_3$ denote hyperparameters that weight the $L_{cls}$, $L_{reg}$, and $L_{con}$, respectively. We empirically set $\lambda_1$, $\lambda_2$, and $\lambda_3$ as 1, 0.001, 0.01.

### 2.4.2 Anatomical Location Identification Task

The aim of this task is not only to inform the location of CMBs but also to eliminate the false positives candidates from the detection process. As described in the data labeling section, there are four distinct categories of microbleeds depending on their anatomical location: lobar, deep, infratentorial, and 'none'. If the detected CMB candidate is located over the 'none' region, it is considered a clear CMB mimic, and we automatically remove it. Unlike FreeSurfer tool, the model of this task can be input not only T1 images but also SWI images. In addition, in the case of FreeSurfer, it takes about 8 hours or more to segment the brain per subject, whereas this model takes about 40 seconds per subject.

#### 2.4.2.1 Network Architecture

Unlike the detection task, either SWI or T1-MPRAGE images, excluding phase images, are used as input into the model because they contain structural information of the brain. We additionally input three tensors which have absolute coordinate information. This is because that the SWI or T1-MPRAGE images are cropped before being input into the model and loses its absolute coordinate information. The three tensors are normalized so that their intensities range from 0 to 1 along the x, y, and z axes. When data cropping is performed, the tensors are also cropped to the same coordinates. By inputting the cropped tensors with cropped image, it would be helpful for the model to infer where the cropped images are cropped from the whole brain image. This task requires performing a segmentation of the four regions. Thus, we employed the original U-Net, which upsamples the generated feature maps into their original size as shown in Figure 5. In the end, there are four output channels for lobar, deep, infratentorial, and 'none' regions. The model is trained using dice loss.

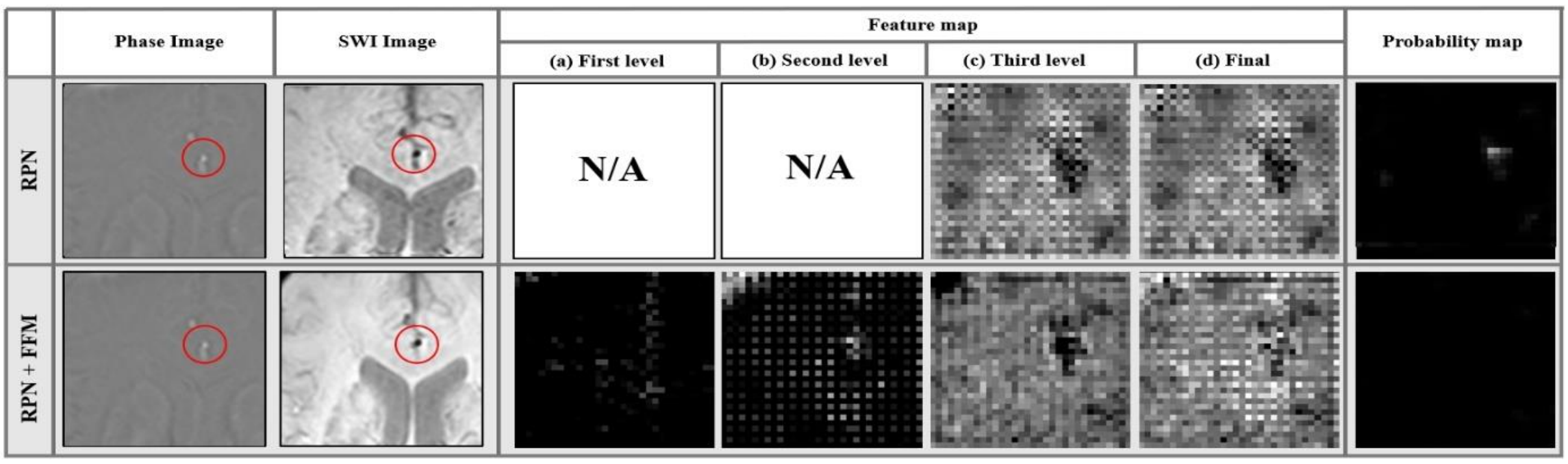

Fig. 6. The feature maps and their generated probability map of the vanilla RPN and RPN with FFM. The lesion in red circle is CMB mimic. (a), (b), and (c) show the feature maps from first to third levels. (d) shows the final feature map. The probability map is also shown in the far right. Note that RPN+FFM reduces the detection probability for this CMB mimic.

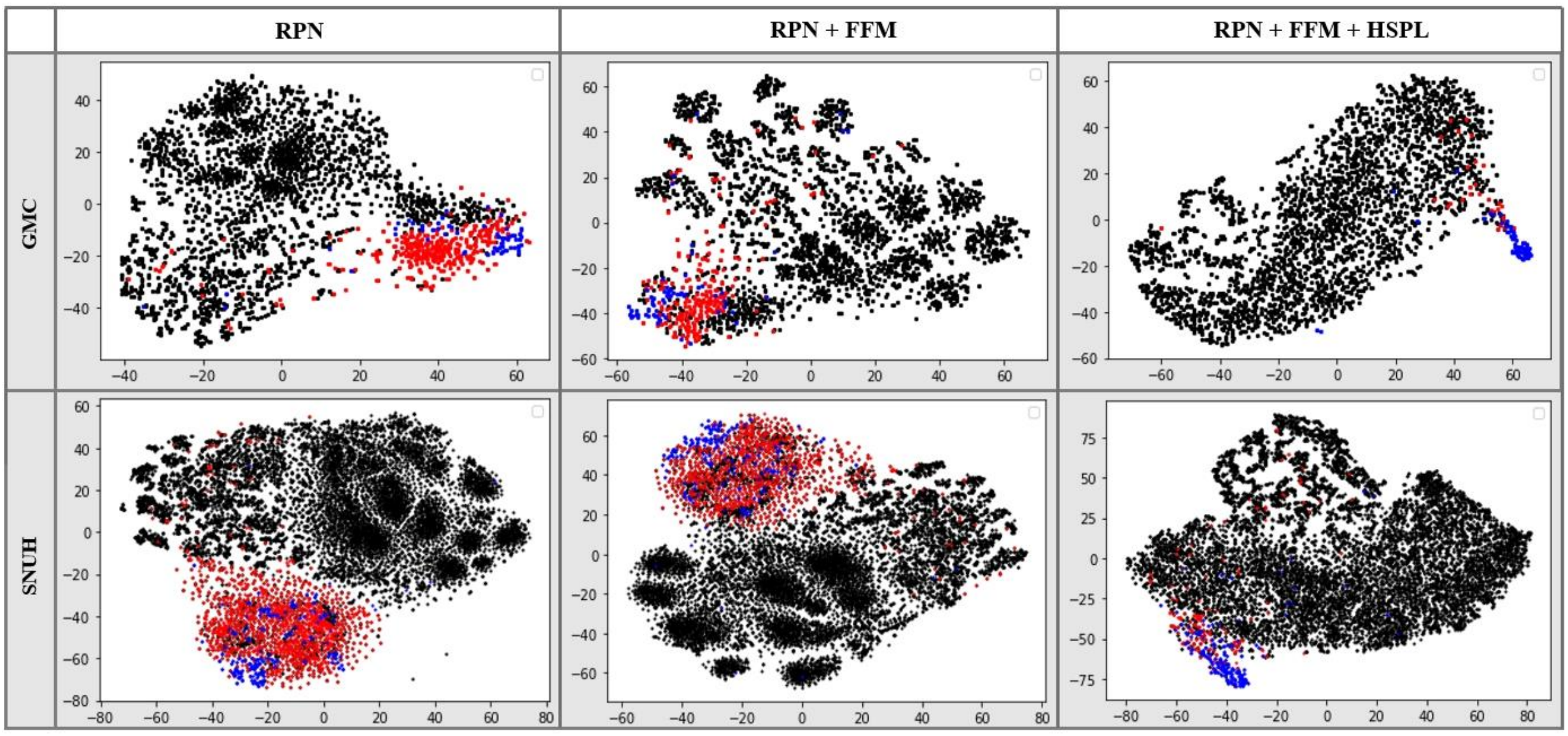

Fig. 7. Feature vectors of CMBs and non-CMBs for three models. The dimension of feature vectors is reduced to two dimensions using t-SNE. The blue, red, black points indicate feature vectors of CMBs, false positives and true negatives, respectively. The top row represents an instance from GMC dataset, while the bottom row shows an example from SNUH dataset. The CMBs (blue) are well separated with RPN+FFM+HSPL approach.

### 2.5 Evaluation Measures

We utilized six evaluation metrics to quantitatively evaluate the capability of the proposed deep 3D deep learning framework for CMB detection and anatomical localization. Sensitivity, precision, and the number of false positives per subject ($FP_{avg}$) are used to evaluate the detection task. Dice score, localization accuracy (LA), and the number of eliminated false positives per subject are utilized for the anatomical localization task evaluation. The evaluation measures are mathematically defined as follows:

$$LA\ (Total) = \frac{TP_{Lobar} + TP_{Deep} + TP_{Infra}}{N_{Lobar} + N_{Deep} + N_{Infra}} \tag{5}$$

$$LA\ (Lobar) = \frac{TP_{Lobar}}{N_{Lobar}}, LA\ (Deep) = \frac{TP_{Deep}}{N_{Deep}}, LA\ (Infra) = \frac{TP_{Infra}}{N_{Infra}} \tag{6}$$

$$EFP_{avg} = \frac{EFP_{CMB}}{N_s} \tag{7}$$

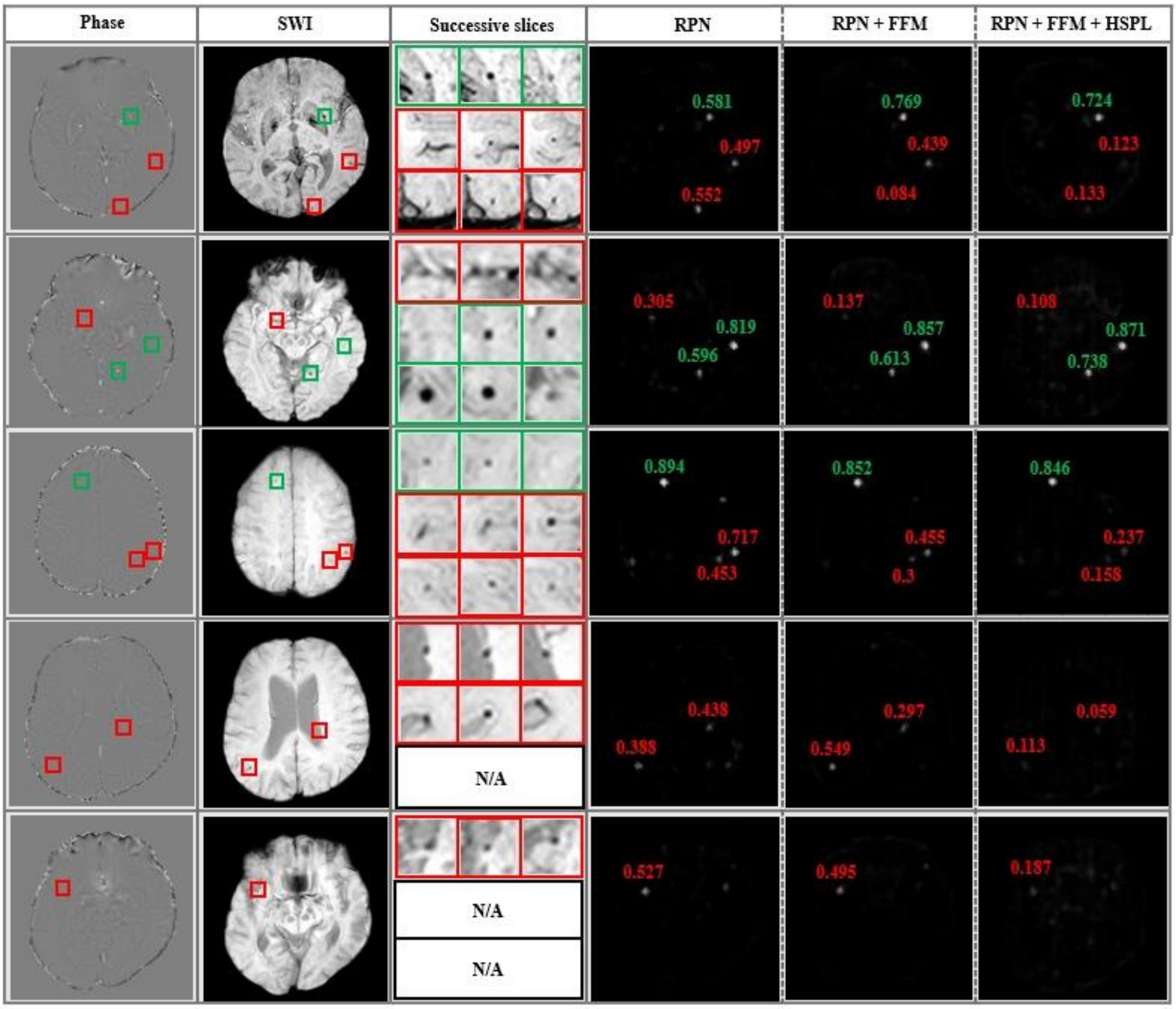

Fig. 8. Examples of the detected candidates using different methods: RPN, RPN with FFM, and RPN with FFM and HSPL. The lesions in green boxes are CMBs and lesions in red boxes are CMB mimics. The values written over the probability maps indicate the probability scores.

where $TP_{Lobar}$, $TP_{Deep}$, and $TP_{Infra}$ denote true positive cases of each class in the anatomisdwcal localization task. The $N_{Lobar}$, $N_{Deep}$, and $N_{Infra}$ indicate the total number of samples of each class. The $EFP_{cmb}$ denotes the number of eliminated false positives by anatomical localization task. $N_s$ indicates the number of subjects.

# 3 Results and Discussion

## 3.1 FFM Analysis

We qualitatively validated the effect of FFM on CMB detection in Figure 6. To evaluate the feature map at each level visually, the channels of feature maps at each level are set to 1. Since the vanilla RPN and the RPN with FFM were trained separately, it is difficult to strictly compare the feature maps of the two models. However, we trained both models in the same condition, and the contrast difference in the final feature map affects the probability map in both models. For vanilla RPN, the feature map of the third level becomes the final feature map. As it is shown in the probability map of vanilla RPN, the regions where the CMB mimic is located show a high probability score. In the case of the RPN with FFM, the final feature map is computed as an aggregation of the feature maps of three levels. As shown in Figure 6 (a) and (b) of the RPN with FFM, they generate regions where the CMB mimics might exist utilizing contextual information. They are added with Figure 6 (c) of the RPN with FFM, which incorrectly predicted the CMB mimic regions as the CMB regions, to produce a corrected Figure 6 (d). Therefore, the probability scores of the RPN with FFM in the regions of CMB mimic get reduced compared to the vanilla RPN case.

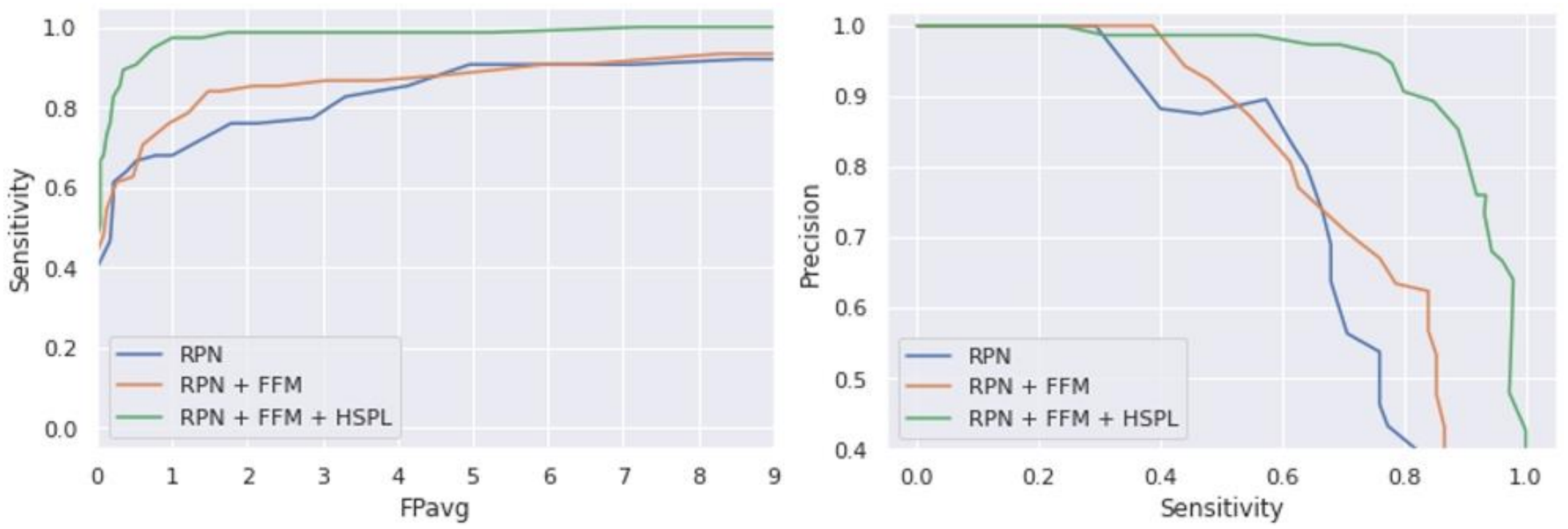

Fig. 9. The left plot shows the sensitivity vs. $FP_{avg}$, while the right plot presents the PR curve for vanilla RPN, RPN with FFM, and RPN with FFM and HSPL.

### 3.2 HSPL Analysis

This section verifies how well the RPN with HSPL could separate the generated features of CMBs and non-CMBs in the feature space. As shown in Figure 7, we plotted dimension-reduced feature vectors of CMBs, false positives, and true negatives. We randomly extracted feature vectors of true negatives from regions not close to the regions of CMBs and false positives. In the case of not including the HSPL, the feature vectors of the CMBs are not aggregated among themselves and are mixed with the feature vectors of the non-CMBs. On the other hand, when HSPL is applied, the feature vectors of CMBs become close to each other and are mixed with fewer non-CMBs. This figure also demonstrates that the proposed HSPL could significantly reduce the number of false positives (red dots).

For the quantitative evaluation, we compared the probability maps of the vanilla RPN, RPN with FFM, and RPN with FFM and HSPL. As shown in Figure 8, we observed that the shape and position of the CMBs do not change within the successive slices compared to the cases of CMB mimics. For this reason, we trained the network using 3D information. The results show that the inclusion of FFM enables the network for better differentiation between true CMBs and CMB mimics. However, significant reduction of FPs was accomplished through using the proposed HSPL as clearly shown in this figure.

Figure 9 presents the sensitivity vs. $FP_{avg}$ and PR plots for the vanilla RPN, RPN with FFM, and RPN with FFM and HSPL. Obviously, the RPN with FFM and HSPL detected the CMBs with a true positive rate of greater than 0.95 at less than 1 $FP_{avg}$. The PR curve in Figure 9 shows that the proposed network with HSPL achieved the highest AUC-PR of 0.94 compared to the vanilla RPN and RPN with FFM of 0.75 and 0.79, respectively. These results prove the ability of the proposed HSPL in reducing the FPs.

Table 1. Performance of various detectors using two different datasets: GMC and SNUH

| Method (# of stage) | | GMC dataset | | | | SNUH dataset | | |
|---|---|---|---|---|---|---|---|---|
| | | Subjects with CMBs | | | Normal | Subjects with CMBs | | |
| | | Sensitivity (%) | Precision (%) | $FP_{avg}$ | $FP_{avg}$ | Sensitivity | Precision | $FP_{avg}$ |
| Al-masni et al., 2020 (2) | 1st: 2D YOLO-v2 | 93.33 | 5.36 | 53.69 | 57.92 | 92.60 | 5.01 | 57.85 |
| | 2nd: 3D CNN | 94.28 | 61.11 | 1.75 | 1.85 | 93.75 | 55.78 | 2.27 |
| | 1st + 2nd | 88.00 | 61.11 | 1.75 | 1.85 | 86.81 | 55.78 | 2.27 |
| 2D EfficientDet-D3 (1) | | 94.66 | 10.33 | 26.39 | 30.50 | 92.52 | 8.78 | 24.36 |
| 3D RPN (1) | | 93.33 | 17.11 | 14.73 | 12.07 | **94.21** | 16.83 | 15.39 |
| 3D RPN + FFM (1) | | 93.33 | 25.64 | 8.82 | 9.00 | 93.89 | 22.70 | 10.57 |
| 3D RPN + FFM + HSPL (1) | | **94.66** | **78.02** | **0.86** | **0.85** | 93.89 | **60.70** | **2.01** |

Furthermore, we compared the performance of a two-stage framework developed by Al-masni et al. [26], EfficientDet-D3 [44], and vanilla RPN, RPN with FFM, and RPN with FFM and HSPL on both GMC and SNUH datasets. All these models were trained using 91 subjects containing 290 CMBs from GMC dataset and tested on 37 subjects containing 75 CMBs from GMC dataset, and 94 subjects containing 311 CMBs from SNUH dataset.

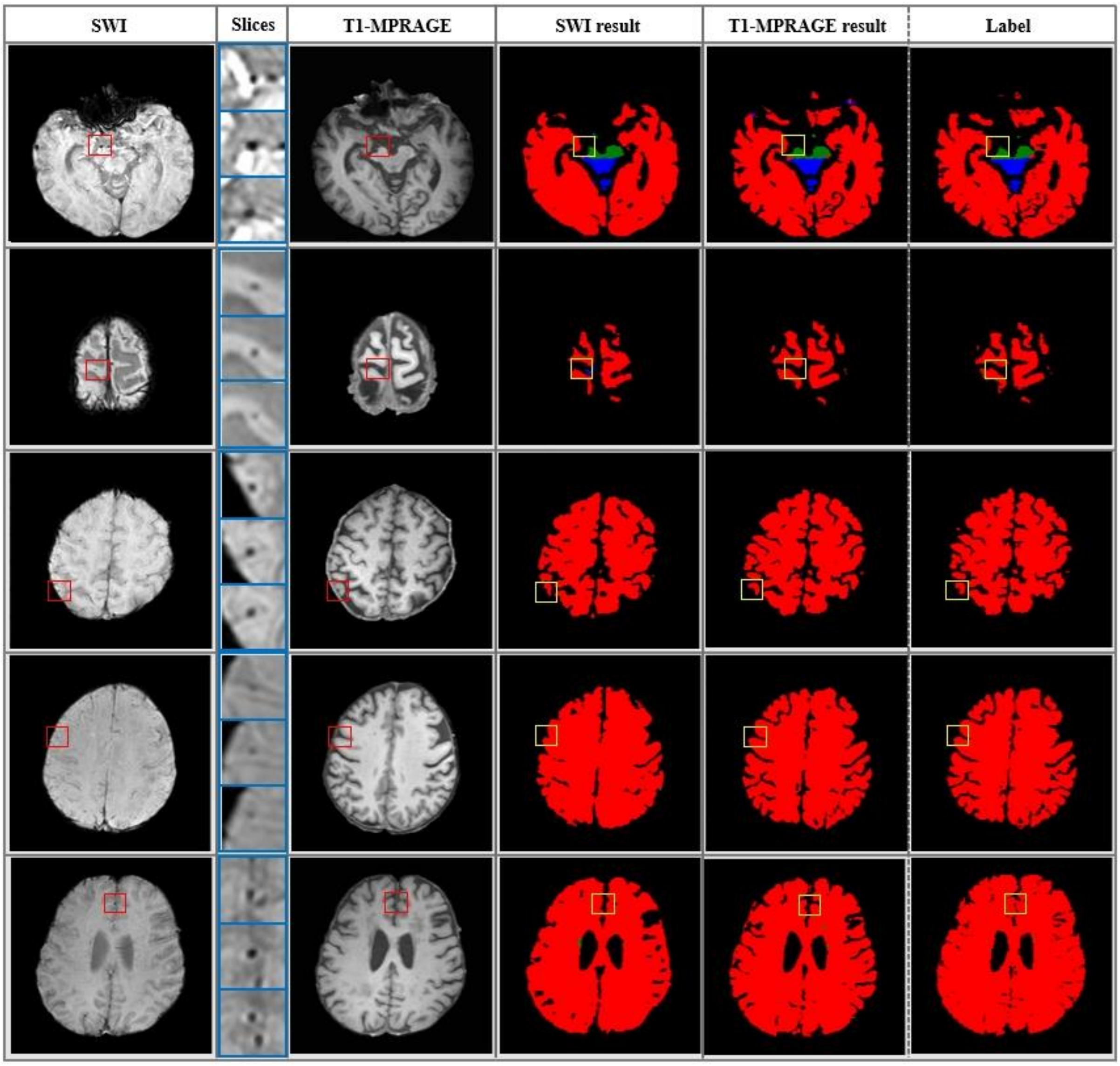

Fig. 10. The lesions in boxes are CMB candidates from detection task. The red, green, and blue regions indicate lobar, deep, and infratentorial, respectively. All these detected CMB candidates get eliminated after checking the segmentation results where the candidates exist out of the anatomical regions.

All models are conducted using a batch size of 1, learning rate of 0.01, momentum of 0.9, and step learning rate scheduler's step size and gamma of 50 and 0.5, respectively. As shown in Table 1, when 3D RPN is compared against 2D detection models, the 3D RPN shows a precision of 17.11%, which is 11.75% higher than 2D YOLO-v2 and 6.78% higher than 2D EfficientDet-D3 while maintaining similar sensitivity.[44]. This result implies that 3D information is very important for CMB detection. To demonstrate the ability to capture contextual information of FFM and efficiency to concentrate hard sample of HSPL for CMB mimics reduction, vanilla 3D RPN, 3D RPN with FFM, and 3D RPN with FFM and HSPL were compared by setting each with an optimal hyperparameter. When FFM was added, $FP_{avg}$ was reduced by half, and especially, when HSPL was added, the performance of 3D RPN was improved with major improvements of 60.91% in Precision and 95.38% in $FP_{avg}$. Our single-stage detector, 3D RPN with FFM and HSPL, shows better performance than the two-stage detector that utilized YOLO-v2 and 3D-CNN [26]. This result shows that the HSPL can replace the role of the second stage (i.e., classification model) that intensively distinguishes CMBs and CMB mimics. In actual clinical setting, since there are more subjects without CMBs than subjects with CMBs, it is very important that the number of false positives should be low in subjects without CMBs. For this reason, we also tested the 14 normal subjects, and the results recorded $FP_{avg}$ similar to those of subjects with CMBs. In the case of SNUH dataset, which is utilized as a generalization test, it seems that the proposed detection method is somewhat well generalized, although the performance decreased slightly due to the different MRI vendors and MRI parameters.

### 3.3 Performance of Anatomical Localization Task

Table 2. Performance of anatomical localization task on three types of data.

| | GMC dataset | | | | | | SNUH dataset | |
|---|---|---|---|---|---|---|---|---|
| | SWI | | | T1-MPRAGE | | | SWI | |
| | Dice score | LA | $EFP_{avg}$ | Dice score | LA | $EFP_{avg}$ | LA | $EFP_{avg}$ |
| Lobar | 0.9043 | 0.97 | N/A | 0.9491 | 1.0 | N/A | 0.97 | N/A |
| Deep | 0.8544 | 1.0 | N/A | 0.8846 | 0.91 | N/A | 0.95 | N/A |
| Infratentorial | 0.9251 | 1.0 | N/A | 0.9518 | 1.0 | N/A | 0.97 | N/A |
| Total | 0.8946 | 0.98 | 0.30 | 0.9272 | 0.98 | 0.34 | 0.97 | 0.28 |

The performance of the anatomical localization task using a dice score that measures the segmentation performance, a localization accuracy that measures how well the anatomical location of the CMBs is classified, and $EFP_{avg}$ that measures how many false positives from the detection task are eliminated is shown in Table 2. The dice score evaluation divides the regions into the four classes: lobar, deep, infratentorial, and CMBs cannot be exist. This evaluation shows how well the model classifies the exact location of CMBs and reduces false positives. Comparing the performance when utilizing T1-MPRAGE and SWI of GMC, the T1-MPRAGE image outperforms the SWI image in all evaluations on dice score because the anatomical contrast of T1-MPRAGE image is better than SWI image. In the case of SNUH dataset, the label acquisition through FreeSurfer was not possible because the T1-MPRAGE image did not exist in SNUH dataset. Therefore, the neuroradiologist labeled the anatomical location of CMBs manually and the model of anatomical localization task calculated localization accuracy and the $EFP_{avg}$. Both evaluations show similar results to GMC dataset.

As shown in Figure 10, we observed that the shape and position of the CMB candidates do not change within the successive slices. For this reason, the detection task classified the CMB candidates as CMBs. However, these CMB candidates are located in regions where CMB could not exist anatomically. So, the anatomical localization task eliminated the CMB candidates easily. It is of note that the segmentation results are more accurate when T1-MPRAGE images are used as input compared to SWI images.

### 3.4 Performance of Overall Framework

Table 3. Performance of whole framework.

| Task (Segmentation input data) | GMC dataset | | | | SNUH dataset | | |
|---|---|---|---|---|---|---|---|
| | Subjects with CMBs | | | Normal | Subjects with CMBs | | |
| | Sensitivity (%) | Precision (%) | $FP_{avg}$ | $FP_{avg}$ | Sensitivity (%) | Precision (%) | $FP_{avg}$ |
| Detection | 94.66 | 78.02 | 0.86 | 0.85 | **93.89** | 60.70 | 2.01 |
| Detection + Anatomical localization (SWI) | 94.66 | 84.52 | 0.56 | 0.64 | 92.60 | **67.92** | **1.44** |
| Detection + Anatomical localization (T1) | **94.66** | **85.54** | **0.52** | **0.57** | N/A | N/A | N/A |

Table 3 shows the performance of the whole framework. As shown in Table 3, for subjects with CMBs of GMC dataset, there are no cases of misclassifying true positives from the detection task as none class in the anatomical localization task when SWI or T1-MPRAGE are used as input. So, the sensitivity is preserved. On the other hand, for the false positives from the detection task, seven false positives when using SWI and eight false positives when using T1-MPRAGE were classified as none class indicating regions where CMBs could not exist and correctly eliminated. In the case of normal subjects, three false positives when using SWI and four false positives when using T1-MPRAGE were removed. In both subjects with CMBs and normal subjects, the $FP_{avg}$ decreased a little more when T1-MPRAGE was used as input compared to the case of using SWI.

In the case of SNUH dataset, the four true positives from the detection task are eliminated by misclassifying the anatomical location of true positives into none class in the anatomical localization task. However, the 53 false positives from the detection task are eliminated and the precision increased from 60.70% to 67.92% in the anatomical localization task.

### 3.5 Limitation and Future Work

The HSPL randomly crops the data to obtain the non-CMB samples from the entire MR brain image; however, the point with the highest confidence score in the cropped data is considered where the CMB mimic is located. For this reason (i.e., randomly cropped non-CMBs), there is a limitation where the results are varied for each training, which could lead to low reproducibility. To solve this to some extent, enough epochs are set so that the cropped data cover all regions of the entire brain image considering the ratio between the size of the cropped data and the size of the entire brain image.

Further, due to the property of SWI, it is more ambiguous to distinguish the brain structurally than T1-MPRAGE, so the performance of anatomical segmentation is lower. For future work, utilizing cross-modality translation from the SWI to the T1-MPRAGE, we will increase the performance when using SWI as much as the performance when T1-MPRAGE is used as input.

## 4 Conclusion

In this paper, we present a framework that not only detects CMBs in the detection task but also informs the exact anatomical location of CMBs in the newly proposed anatomical localization task. In the case of the detection task, we proposed the Feature Fusion Module (FFM) that reduces false positives by injecting contextual information into the final feature map, and the Hard Sample Prototype Learning (HSPL) that allows the model to concentrate on the CMB mimics (i.e., hard samples). The proposed modules in our single-stage detector outperform the two-stage detectors without using any classification model. Further, the main purpose of the anatomical localization task is to inform the anatomical location of CMBs, which leads to identifying and eliminating false positives in the regions where CMBs could not exist, decreasing the $FP_{avg}$.

**Acknowledgements.** This research was supported by the Brain Research Program through the National Research Foundation of Korea (NRF) funded by the Ministry of Science, ICT & Future Planning (2018M3C7A1056884) and (NRF - 2019R1A2C1090635), Korea Healthcare Technology R&D Project through the Korea Health Industry Development Institute (KHIDI) (HI14C1135), Korea Medical Device Development Fund grant funded by the Korea government (the Ministry of Science and ICT, the Ministry of Trade, Industry and Energy, the Ministry of Health & Welfare, Republic of Korea, the Ministry of Food and Drug Safety) (Project Number: 202011D23)